\documentclass[english,reprint,aps,prb,superscriptaddress]{revtex4-2}
\usepackage[T1]{fontenc}
\usepackage[utf8]{inputenc}
\usepackage{amsmath}
\usepackage{amssymb}
\usepackage{graphicx}
\usepackage{textcomp}

\makeatletter
\usepackage{hyperref}
\usepackage{color}
\usepackage{siunitx}

\makeatother

\usepackage{babel}
\begin{document}
\title{Revealing the anisotropic charge-density-wave order of TiSe$_2$ through high harmonic generation}

\author{Lin Zhang}
\email{lin.zhang@icfo.eu}
\affiliation{ICFO-Institut de Ciencies Fotoniques, The Barcelona Institute of Science and Technology, Av. Carl Friedrich Gauss 3, 08860 Castelldefels (Barcelona), Spain}

\author{Igor Tyulnev}
\affiliation{ICFO-Institut de Ciencies Fotoniques, The Barcelona Institute of Science and Technology, Av. Carl Friedrich Gauss 3, 08860 Castelldefels (Barcelona), Spain}

\author{Lenard Vamos}
\affiliation{ICFO-Institut de Ciencies Fotoniques, The Barcelona Institute of Science and Technology, Av. Carl Friedrich Gauss 3, 08860 Castelldefels (Barcelona), Spain}

\author{Julita Poborska}
\affiliation{ICFO-Institut de Ciencies Fotoniques, The Barcelona Institute of Science and Technology, Av. Carl Friedrich Gauss 3, 08860 Castelldefels (Barcelona), Spain}

\author{Utso Bhattacharya}
\affiliation{Institute for Theoretical Physics, ETH Zurich, Zurich, Switzerland}

\author{Ravindra W. Chhajlany}
\affiliation{ISQI - Institute of Spintronics and Quantum Information, Faculty of Physics and Astronomy, Adam Mickiewicz University, Pozna{\' n}, Poland}

\author{Tobias Grass}
\affiliation{DIPC - Donostia International Physics Center, Paseo Manuel de Lardiz{\' a}bal 4, San Sebasti{\' a}n, Spain}
\affiliation{Ikerbasque - Basque Foundation for Science, Maria Diaz de Haro 3, Bilbao, Spain}

\author{Jens Biegert}
\affiliation{ICFO-Institut de Ciencies Fotoniques, The Barcelona Institute of Science and Technology, Av. Carl Friedrich Gauss 3, 08860 Castelldefels (Barcelona), Spain}
\affiliation{ICREA, Pg. Llu{\' i}s Companys 23, Barcelona, Spain}

\author{Maciej Lewenstein}
\affiliation{ICFO-Institut de Ciencies Fotoniques, The Barcelona Institute of Science and Technology, Av. Carl Friedrich Gauss 3, 08860 Castelldefels (Barcelona), Spain}
\affiliation{ICREA, Pg. Llu{\' i}s Companys 23, 08010 Barcelona, Spain}

\begin{abstract}
Titanium diselenide (TiSe$_{2}$) is a transition-metal dichalcogenide material that undergoes a charge-density-wave (CDW) transition at $T_{c}\approx 200\,\mathrm{K}$. In a recent experiment [I. Tyulnev {\it et al.}, Commun. Mater. 6, 152 (2025)], the high harmonic generation (HHG) spectra of this material has been studied, which exhibits asymmetric behavior with respect to the polarization angle of the incident light and provides a new perspective to the CDW phase transition. In this work, we work out a theoretical explanation for the experimentally observed behavior of HHG spectra. We propose a simplified phenomenological mean-field model for this material, based on which the HHG spectra is calculated through the time-dependent Schr{\" o}dinger equation. This model correctly describes the measured intensity distribution of the third-, fifth-, and seventh-order harmonic generation as a function of polarization direction and reveals a strong asymmetry due to the anisotropic CDW order in the low temperature phase. Our work provides a basis for applying high harmonic spectroscopy to reveal a new perspective on the nature of CDW orders.
\end{abstract}
\maketitle

\section{Introduction}

High harmonic generation (HHG) is a nonlinear and nonperturbative optical process that occurs when intense laser fields interact with matter~\cite{Corkum1993,Lewenstein1994}. This process leads to the ultrafast emission of high energy photons with frequencies being integer multiples of the initial driving frequency. HHG was initially discovered in the atomic and molecular gases and is the basis of attosecond science~\cite{Krausz2009,Schultz2014}. The recent observation of HHG from solids has further renewed the interest in this field~\cite{Ghimire2011,Schubert2014,Hohenleutner2015,Luu2015,Vampa2015,Ndabashimiye2016,Liu2016,You2016,Yoshikawa2017,Ghimire2018}, which provides ultrafast time-resolved measurement about the driven material properties, such as the electron dynamics~\cite{Langer2016,Reimann2018,Baudisch2018,Uzan2020}, the electron band properties~\cite{Vampa2015a,Lanin2017,Luu2018,Lv2021}, symmetries~\cite{Langer2017,Langer2018,Neufeld2019,Heinrich2021,Yue2022a}, and so on. 

Theoretically, the HHG in solids is understood via the semiclassical single-active-electron three-step model~\cite{Corkum1993,Lewenstein1994} adapted from the atomic HHG by assuming weak correlations or an effective single-particle picture, in which the electrons traverse the valence and conduction bands, creating and recombining with holes, producing a harmonic spectrum~\cite{Ghimire2018,Vampa2014,Huttner2017,Kruchinin2018,Yu2019,Yue2022}. Replacing these fundamental electrons with the many-body excitations such as the doublons and holons, the HHG can also be applied to study the properties of strongly correlated systems, such as the Mott insulator~\cite{Silva2018,Tancogne-Dejean2018,Orthodoxou2021,Murakami2021,Murakami2022,Uchida2022}, Kitaev spin liquid~\cite{Kanega2021}, and high-temperature superconductor~\cite{Alcala2022}, in which the HHG spectra could be sensitive to the underlying different many-body phases and provides an all-optical probe of the quantum phase transitions~\cite{Alcala2022,Shao2022}.

As a layered transition-metal dichalcogenide material, the titanium diselenide (TiSe$_{2}$) undergoes a transition from the high temperature semimetal or semiconductor phase with lattice constant $a=3.54\,\mathring{\mathrm{A}}$ into a commensurate $2\times2\times2$ charge-density-wave (CDW) state at transition temperature $T_{c}\approx 200\,$K~\cite{DiSalvo1976}. Moreover, this material exhibits superconductivity with melting of the CDW upon the copper intercalation~\cite{Morosan2006,Li2007a}, applying pressures~\cite{Kusmartseva2009,Joe2014}, or carrier doping~\cite{Li2015,Luo2016}. There is ongoing debate over the underlying mechanism, i.e., whether the CDW instability is driven by the electron-phonon coupling (Jahn-Teller effect) or exciton condensation~\cite{Kidd2002,Rossnagel2002,Cercellier2007,Wezel2010,Li2023}. Therefore, clarifying the properties of CDW order is significant for understanding this material. The very recent experiment of Ref.~\cite{Tyulnev2024} shows that the HHG is highly sensitive to the CDW order and exhibits pronounced asymmetric behavior with respect to the polarization angle of the incident light. This provides a new perspective for studying the CDW phase transition.

In this work, we provide a theoretical explanation for the experimentally observed behavior of HHG spectra. Specifically, a simplified phenomenological mean-field model is proposed for the CDW phase, based on which the HHG spectra as a function of laser polarization angle are calculated. Our model correctly describes the measured behavior of harmonic orders as a function of the driving field polarization angle. The polarimetry sensitively detects the crystal orientation and explains the single-peaked behavior of the third-order (H3) and seventh-order (H7) harmonic generation as well as the double-peaked behavior of the fifth-order harmonic generation (H5). Based on the excellent match between the experiment and theory, we reveal a strong asymmetry due to the anisotropic CDW order in the low temperature phase of the material. Our work theoretically explains the asymmetric behavior of HHG spectra observed in Ref.~\cite{Tyulnev2024} and unveils that the HHG is sensitive to the CDW order.

\section{Mean-field model}

\begin{figure}
\includegraphics{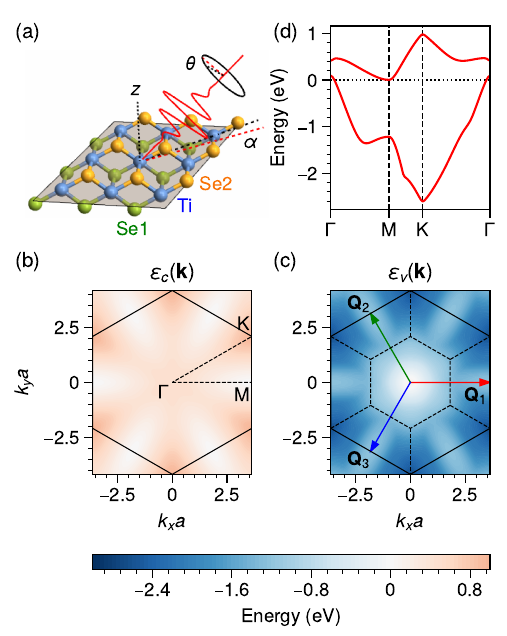}
\caption{Setup and tight-binding band structure of $\mathrm{TiSe}_{2}$. (a) Laser geometric configuration. The laser pulse (solid red line) with linear polarization angle $\theta$ hits the material with an incident angle with respect to the material plane or the $z$ axis, and the crystal axis (in-plane black dashed line) has an offset angle $\alpha$ compared to the in-plane incidence direction of the laser field (in-plane red dashed line).   (b) (c) Contour plots of the lowest lying conduction band $\varepsilon_{c}(\mathbf{k})$ and the highest lying valence band $\varepsilon_{v}(\mathbf{k})$. The solid black hexagons represent the first Brillouin zone. In (c), we also show the wave vectors $\mathbf{Q}_{1,2,3}$ of the CDW order, which couple the hole pocket of valence band to the electron pockets of conduction band and fold the Brillouin zone. The dashed black lines represent the reduced Brillouin zone. (d) Band structure along the high-symmetry line [cf.~(b)]. Here the Fermi energy $E_{F}$ is set to zero.\label{fig:figure1}}
\end{figure}

We consider a simplified phenomenological mean-field model to describe the CDW order of TiSe$_{2}$. As the layers making up TiSe$_2$ are separated from each other by a van der Waals gap and the laser field mainly interacts with the surface layer, we model the material via a two-dimensional Hamiltonian.  Like the exciton condensate model~\cite{Jerome1967,Monney2009}, we consider only the highest lying valence band and the lowest lying conduction band within the minimal model for simplicity. The total Hamiltonian is expressed as 
\begin{equation}\label{eq:mean-field model}
\begin{aligned}H & =\sum_{\mathbf{k}}[\varepsilon_{v}(\mathbf{k})-\mu]a_{\mathbf{k}}^{\dag}a_{\mathbf{k}}+\sum_{\mathbf{k}}[\varepsilon_{c}(\mathbf{k})-\mu]b_{\mathbf{k}}^{\dag}b_{\mathbf{k}}\\
 & \quad+\sum_{\mathbf{Q},\mathbf{k}}[\Delta_{(-\mathbf{Q})}a_{\mathbf{k}+\mathbf{Q}}^{\dag}b_{\mathbf{k}}+\Delta_{(+\mathbf{Q})}b_{\mathbf{k}-\mathbf{Q}}^{\dag}a_{\mathbf{k}}].
\end{aligned}
\end{equation}
Here $a_{\mathbf{k}}$ and $b_{\mathbf{k}}$ are the annihilation operator for the valence band and conduction band, respectively. The dispersions of valence band $\varepsilon_{v}(\mathbf{k})$ and conduction band $\varepsilon_{c}(\mathbf{k})$ are obtained from the tight-binding model introduced in Ref.~\cite{Kaneko2018} (see also Appendices for more details). As shown in Figs.~\ref{fig:figure1}(b) and \ref{fig:figure1}(c), the valence band composed mainly of Se $4p$ orbitals gives rise to a hole pocket centered as $\Gamma$ point, while the conduction band dominated by Ti $3d$ orbitals has three electron pockets centered at the M points. These pockets exhibit a small band overlap [see Fig.~\ref{fig:figure1}(d)], rendering a semimetal phase in the high-temperature regime of TiSe$_{2}$.

Below the critical temperature $T_{c}=200\,\mathrm{K}$, the electron-electron and electron-phonon interactions further lead to a triple-$\mathbf{Q}$ CDW order with wave vectors $|\mathbf{Q}_{1}|=|\mathbf{Q}_{2}|=|\mathbf{Q}_{3}|=\Gamma\mathrm{M}$, which couple the hole pocket of valence band to the electron pockets of conduction band; see Fig.~\ref{fig:figure1}(c). In our phenomenological mean-field model \eqref{eq:mean-field model}, we model the CDW of TiSe$_{2}$ with the mean-field order parameters $\Delta_{(+\mathbf{Q})}$ while ignore the corresponding microscopic mechanism when studying the HHG spectra of this material. We assume that the CDW order paramter follows the mean-field temperature dependence considered in Ref.~\cite{Monney2011}:
\begin{equation}\label{eq:mean-field CDW order parameter}
    \Delta(T)=\Delta_{0}\sqrt{1-(T/T_{c})^2}
\end{equation}
with $\Delta_{0}=115\,\mathrm{meV}$. The CDW order parameters fold the Brillouin zone; see Fig.~\ref{fig:figure1}(c). In the reduced Brillouin zone (RBZ), the mean-field Hamiltonian can be rewritten as
\begin{equation}
    H=\sum_{\mathbf{k}\in\mathrm{RBZ}}\Psi^{\dagger}_{\mathbf{k}}H(\mathbf{k})\Psi_{\mathbf{k}},
\end{equation}
where $H(\mathbf{k})$ is the $8\times 8$ momentum-dependent Hamiltonian matrix and we have the basis $\Psi_{\mathbf{k}}=(a_{\mathbf{k}}, b_{\mathbf{k}}, a_{\mathbf{k}+\mathbf{Q}_{1}}, b_{\mathbf{k}+\mathbf{Q}_{1}}, a_{\mathbf{k}+\mathbf{Q}_{2}}, b_{\mathbf{k}+\mathbf{Q}_{2}}, a_{\mathbf{k}+\mathbf{Q}_{3}}, b_{\mathbf{k}+\mathbf{Q}_{3}})^{\mathrm{T}}$. 

The above simplified phenomenological mean-field model of course has limitations. For example, the band dispersion $\varepsilon_{c/v}(\mathbf{k})$ from the tight-binding model used in this work only captures the main feature of the more accurate one obtained from density functional theory (DFT) calculation~\cite{Kaneko2018}, which could overestimate or underestimate the HHG intensity at certain polarization angle. Also, the multi-orbital and multi-band effects are ignored in our modeling, which may have nontrivial influence on the HHG spectra. However, as we show below, our simplified phenomenological mean-field model still qualitatively captures the HHG spectra of $\mathrm{TiSe}_{2}$. Hence our modeling is valid.

\section{HHG spectra}

To investigate the high harmonic spectra of the material, we couple the mean-field Hamiltonian $H(\mathbf{k})$ to an intense laser field described by the vector potential $\mathbf{A}(t)$ in dipole approximation through the minimal coupling scheme $\mathbf{k}\to\mathbf{k}(t)=\mathbf{k}+e\mathbf{A}(t)/\hbar$; see Fig.~\ref{fig:figure1}(a). Here $e$ is the fundamental charge, and $\hbar$ is the Planck constant. Using the velocity gauge equation of motion in the Bloch basis~\cite{Moos2020}, we have the following time-dependent Schr{\" o}dinger equation for the density matrix $\rho(\mathbf{k},t)$ at momentum $\mathbf{k}$:
\begin{equation}\label{eq:time-dependent Schrodinger equation}
     \mathrm{i}\hbar\frac{\mathrm{d}}{\mathrm{d}t}\rho(\mathbf{k},t)=[H(\mathbf{k}+e\mathbf{A}(t)/\hbar;\{\Delta_{\mathbf{Q}_{i}}(t)\}),\rho(\mathbf{k},t)].
\end{equation}
Here we explicitly express the dependence of the Bloch Hamiltonian on the time-evolved CDW order parameter $\Delta_{\mathbf{Q}_{i}}(t)=(U_{\mathbf{Q}_{i}}/V)\sum_{\mathbf{k}}\mathrm{Tr}[\rho(\mathbf{k},t)a^{\dagger}_{\mathbf{k}+\mathbf{Q}_{i}}b_{\mathbf{k}}]$, with $V$  the crystal volume and $U_{\mathbf{Q}_{i}}$ the effective interaction strength between electrons. The latter is taken as a phenomenological temperature-dependent parameter in our modeling to obtain suitable strength of the initial CDW order parameters self-consistently.

Initially, the system is in equilibrium, and the Fermi-Dirac distribution describes the occupation of each momentum mode,  $\rho(\mathbf{k},0)=1/[e^{H(\mathbf{k})/k_{\mathbf{B}}T}+1]$. Then, the system evolves according to the above equation of motion. We find it useful to introduce a phenomenological dephasing effect in the numerical simulation of HHG as follows~\cite{Yue2022}: After each time step $\delta t$ of the evolution, we transform the density matrix into an adiabatic basis $U(\mathbf{k},t)$ obtained by diagonalizing the instantaneous Hamiltonian $H(\mathbf{k}+e\mathbf{A}(t)/\hbar;\{\Delta_{\mathbf{Q}_{i}}(t)\})$, i.e., $\Tilde{\rho}(\mathbf{k},t)=U^{\dagger}(\mathbf{k},t)\rho(\mathbf{k},t)U(\mathbf{k},t)$. Then we apply the dephasing as $\tilde{\rho}_{mn}(\mathbf{k},t)\to\tilde{\rho}_{mn}(\mathbf{k},t)e^{-\delta t/\tau}$ for $m\neq n$ with $\tau$ being the dephasing time.  Finally, we transform back to the Bloch basis.

With the time evolved density matrix $\rho(\mathbf{k},t)$, we can calculate the expectation value of the velocity operator in direction $j$ as 
\begin{equation}
v_{j}(\mathbf{k},t)=\mathrm{Tr}[\rho(\mathbf{k},t)\partial_{k_{j}}H(\mathbf{k}+e\mathbf{A}(t)/\hbar;\{\Delta_{\mathbf{Q}_{i}}(t)\})].
\end{equation}
Denoting the direction of the laser field as $\boldsymbol{n}_{\mathbf{A}}$ and $v_{j}(t)=(1/V)\sum_{\mathbf{k}}v_{j}(\mathbf{k},t)$, the high harmonic spectra can be obtained by the Fourier transform of the velocity (i.e., the current):
\begin{equation}
P(\omega)=\omega^{2}|\mathrm{FT}[\boldsymbol{v}\cdot\boldsymbol{n}_{\mathbf{A}}]|^2.
\end{equation}
In the following, we denote the spectrum strength at $n$th harmonic order as H$n$.

We focus on the polarization dependence of the odd harmonic peaks (H3, H5, and H7). The results are organized along our main findings: (i) Unlike the spectra observed in Ref.~\cite{Tyulnev2024}, the harmonic spectra have a six-fold rotational symmetry for the normal-incident laser geometry, where the incident laser field is normal to the material plane. (ii) A better match with the experimental spectra is obtained when the practical $45\text{\textdegree}$-incident laser geometry is considered, where the laser field has an incident angle $45\text{\textdegree}$ with respect to the material plane, similar to the measurement geometry used in experiment. In particular, this choice reproduces a peak splitting characteristic of H5, which is absent in H3 and H7. (iii) The slight asymmetry between the H5 peaks in Ref.~\cite{Tyulnev2024} is further reproduced when a small offset between the crystal axis and the p-polarization of the laser is introduced; see Fig.~\ref{fig:figure1}(a). (iv) To account for the much stronger H5 peak asymmetry experimentally observed in the low-temperature, we break the three-fold rotational symmetry of the mean-field CDW orders and find that tiny deviations from a symmetric mean field can result in a strongly asymmetric harmonic spectrum. In the following, we present the detailed results.

\begin{figure}
\includegraphics{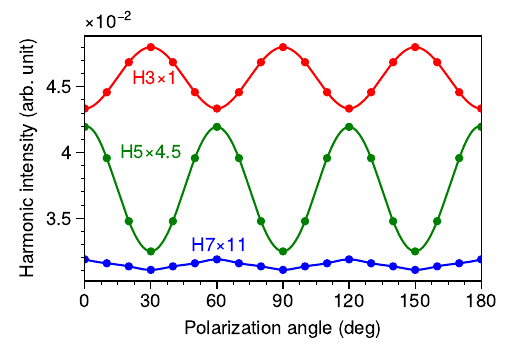}

\caption{HHG spectra as a function of polarization angle for the high temperature semimetal phase in the normal-incident laser geometry. Here we set $A_{0}=1.2\hbar/ea$, $\hbar\omega=0.4\,\mathrm{eV}$, and $n_{\mathrm{cyc}}=8$.}\label{fig:figure2}
\end{figure}

\section{Normal-incident laser geometry}

We first consider the normal-incident laser geometry that can uncover the intrinsic material HHG properties. The incident laser field is normal to the material plane and is given by the vector potential $\mathbf{A}(t)=A(t)[\cos(\theta-60\text{\textdegree}),\sin(\theta-60\text{\textdegree})]$ with polarization angle $\theta$. Here the p-polarized field with $\theta=90\text{\textdegree}$ is along the $\Gamma\mathrm{K}$ direction, and the field strength reads $A(t)=A_{0}\sin^{2}(\omega t/2n_{\mathrm{cyc}})\sin(\omega t)$ with amplitude $A_{0}$, center frequency $\omega$, and cycle number $n_{\mathrm{cyc}}$, which leads to the electric field $E(t)=-\partial A/\partial t=-(\omega/2n_{\mathrm{cyc}})A_{0}\sin(\omega t/n_{\mathrm{cyc}})\sin(\omega t)-\omega A_{0}\sin^{2}(\omega t/2n_{\mathrm{cyc}})\cos(\omega t)$. Solving the time-dependent Schr{\" o}dinger equation \eqref{eq:time-dependent Schrodinger equation} with this vector field leads to the intrinsic material HHG spectra. We set $\hbar\omega=0.4\,\mathrm{eV}$, $n_{\mathrm{cyc}}=8$, and choose $A_{0}=1.2\hbar/ea$ in this work. The resulting electric field with strength $E_{0}=A_{0}\omega\approx 0.14\,\unit{\V/\angstrom}$ is close to the one used in experiment~\cite{Tyulnev2024} with $100\,\unit{\fs}$ duration, central wavelength $3.2\,\unit{\um}$, and field strength $E_{0}\approx 0.08\,\unit{\V/\angstrom}$.

Fig.~\ref{fig:figure2} shows the intrinsic material HHG spectra for the high temperature semimetal phase at $T=300\,\mathrm{K}$ with $\Delta_{\mathbf{Q}_{i}}=0$. The HHG spectra is symmetric with respect to the polarization angle $\theta$ and shows three peaks in each harmonic order at $60\text{\textdegree}$ intervals, reflecting the hexagonal crystal symmetry of TiSe$_2$. As the main contribution to the HHG in the considered high-temperature regime comes from the motion of hole pocket near $\Gamma$ point along the valence band, the observed HHG behavior in Fig.~\ref{fig:figure2} also reflects the different nonlinearity of the valence band dispersion along the $\mathrm{\Gamma K}$ and $\mathrm{\Gamma M}$ direction. This behavior is different from the one observed in the experiment~\cite{Tyulnev2024}, where both H3 and H7 exhibit a single peak for the p-polarized laser field, while H5 splits into two symmetric peaks at $60\text{\textdegree}$ intervals. To reproduce the experimentally observed behavior, we next consider the practical $45\text{\textdegree}$-incident laser geometry used in experiment.

\section{Practical $45\text{\textdegree}$-incident laser geometry}

The HHG spectra in the practical $45\text{\textdegree}$-incident laser geometry are obtained through solving the time-dependent Schr{\" o}dinger equation under the effective in-plane vector field $\mathbf{A}(t)=(A_{x}(t),A_{y}(t))$ with $A_{x}(t)=[\sin\theta\sin(60\text{\textdegree}-\alpha)+(\sqrt{2}/2)\cos\theta\cos(60\text{\textdegree}-\alpha)]A(t)$ and $A_{y}(t)=[\sin\theta\cos(60\text{\textdegree}-\alpha)-(\sqrt{2}/2)\cos\theta\sin(60\text{\textdegree}-\alpha)]A(t)$, which is obtained by projecting the vector potential of the laser field with incident angle $45\text{\textdegree}$ onto the material plane. Here the crystal axis offset $\alpha$ is also introduced in the laser field, describing the situation that the p-polarized field deviates from the $\Gamma\mathrm{K}$ direction; cf. Fig.~\ref{fig:figure1}(a). Compared with the normal-incident laser geometry, the in-plane laser field amplitude now varies with the polarization angle $\theta$, which is maximum when $\theta=90\text{\textdegree}$ and becomes minimized when $\theta=0\text{\textdegree}$. With this, the intrinsic material HHG spectra with three peaks shown in Fig.~\ref{fig:figure2} should be deformed. Particularly, the HHG intensity away from the p-polarization $\theta=90\text{\textdegree}$ should be reduced.

\begin{figure}
\includegraphics{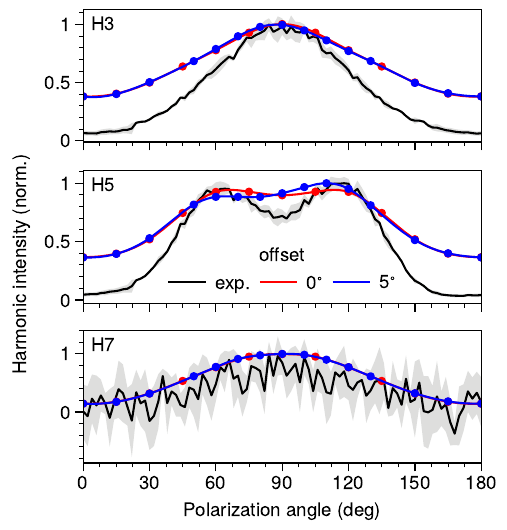}

\caption{Normalized HHG spectra of the high temperature semimetal phase in the practical $45\text{\textdegree}$-incident laser geometry. Here the black lines represent the experimental data in Ref.~\cite{Tyulnev2024} with the shaded area being the error bar. The red lines show the theoretical results without crystal axis offset, while the blue lines are obtained by introducing an offset $\alpha=5\text{\textdegree}$ in the numerical simulation. Here we set $A_{0}=1.2\hbar/ea$, $\hbar\omega=0.4\,\mathrm{eV}$, and $n_{\mathrm{cyc}}=8$. \label{fig:HHG_T_300_with_and_without_offset}}
\end{figure}

In Fig.~\ref{fig:HHG_T_300_with_and_without_offset}, we show the HHG spectra normalized by its maximal value for the high-temperature semimetal phase in the practical $45\text{\textdegree}$-incident laser geometry. The high harmonic H3 and H7 now exhibit a single peak around the polarization angle $\theta=90\text{\textdegree}$, while H5 splits into two symmetric peaks at $\theta=30\text{\textdegree}$ and $\theta=120\text{\textdegree}$ with interval $60\text{\textdegree}$. This is close to the experimental results reported in Ref.~\cite{Tyulnev2024}; cf. the black lines in Fig.~\ref{fig:HHG_T_300_with_and_without_offset}. Hence the experimentally observed peak behavior is a collaborative consequence of the intrinsic
material HHG properties uncovered in the normal-incident laser geometry (cf. Fig.~\ref{fig:figure2}) and the practical laser geometry with incident angle $45\text{\textdegree}$. Especially, while the single peak behavior at polarization angle $90\text{\textdegree}$ in the practical $45\text{\textdegree}$-incident laser geometry could correspond to both the situations with either a peak (H3) or a valley (H7) at $\theta=90\text{\textdegree}$ in the normal-incident laser geometry, the double peak behavior observed in H5 in the practical $45\text{\textdegree}$-incident laser geometry must indicate a valley at $\theta=90\text{\textdegree}$ in the normal-incident laser geometry. We note that in Fig.~\ref{fig:HHG_T_300_with_and_without_offset}, the calculated intensity variations in H3 and H5 over different polarization angles are not as pronounced as in the experiment for the limitation of our model mentioned above and the possible reason that the actual laser incident angle in practical experiment may be slightly different from $45\text{\textdegree}$. However, our results still capture the main physics of HHG spectra.

We note that in the experiment, a slight asymmetric behavior between the peaks in H5 is also observed, which must arise from external factors related to the laser field or affecting the static properties of material that can be effectively captured by the asymmetric band dispersions in our mean-field
description. To explain this observation, we notice that the input laser field in the experiment has an offset angle $7\text{\textdegree}\pm 2\text{\textdegree}$ with respect to the crystal axis; cf. Fig.~\ref{fig:figure1}(a). After introducing an offset $\alpha=5\text{\textdegree}$ in our numerical simulation, the HHG spectra fits very well with the experimental results; see the blue curves in Fig.~\ref{fig:HHG_T_300_with_and_without_offset}. Particularly, the left peak in H5 is lower than the right peak due to this offset. Therefore, the crystal axis offset could be responsible for the asymmetric behavior in the high-temperature phase. 

In general, the nonzero ellipticity of laser field could also induce the asymmetric behavior. However, the laser ellipticity in experiment is too small to lead to visible difference from the linear polarization results hence is ignored. The possible effective asymmetric band dispersions in our mean-field description due to the external factors like disorder, residual strain, surface inhomogeneity and so on could also lead to the asymmetric HHG spectra. However, these factors are not necessary for explaining the high-temperature asymmetric HHG spectra as the HHG asymmetry in the high temperature is quite weak and can be described solely by the crystal axis offset angle. Moreover, introducing these factors could also not explain the low-temperature behavior observed in experiment, as we discuss below. Hence we will not discuss these factors in detail.

\begin{figure}
\includegraphics{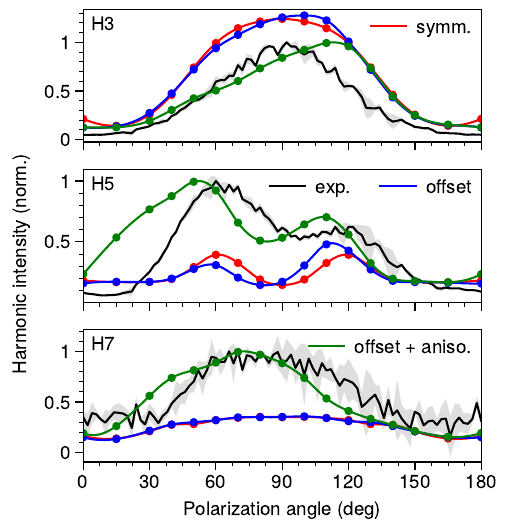}

\caption{Normalized HHG spectra of the low temperature CDW phase in the practical $45\text{\textdegree}$-incident laser geometry. The black lines are the experimental results, while the colored lines are theoretical results. The red lines show the symmetric case with $\Delta_{\mathbf{Q}_{1}}=\Delta_{\mathbf{Q}_{2}}=\Delta_{\mathbf{Q}_{3}}=0.1147\,\mathrm{eV}$, and the blue lines are obtained by introducing a crystal axis offset $\alpha=5\text{\textdegree}$. For the green lines, in addition to the offset $\alpha=5\text{\textdegree}$, we also introduced the anisotropic CDW order parameters $\Delta_{\mathbf{Q}_{1}}=0.1148\,\mathrm{eV}$, $\Delta_{\mathbf{Q}_{2}}=0.1147\,\mathrm{eV}$, and $\Delta_{\mathbf{Q}_{3}}=0.1146\,\mathrm{eV}$. The parameters for the laser pulse are the same as in Fig.~\ref{fig:HHG_T_300_with_and_without_offset}.\label{fig:HHG_T_14}}
\end{figure}

We next consider the low-temperature CDW phase at $14\,\mathrm{K}$. In the absence of strain or disorder, the triple-$\mathbf{Q}$ CDW order parameters in TiSe$_2$ are isotropic due to the three-fold rotation crystal symmetry~\cite{Kaneko2018}. For this, we first consider the CDW order parameters with $\Delta_{\mathbf{Q}_{1}}=\Delta_{\mathbf{Q}_{2}}=\Delta_{\mathbf{Q}_{3}}$. According to the mean-field CDW order parameter \eqref{eq:mean-field CDW order parameter} considered in this work, we have the phenomenological effective interaction strength $U_{\mathbf{Q}_{1}}=U_{\mathbf{Q}_{2}}=U_{\mathbf{Q}_{3}}=-1.80935\,\mathrm{eV}$ and the self-consistent CDW order parameter $\Delta_{\mathbf{Q}_{1}}=\Delta_{\mathbf{Q}_{2}}=\Delta_{\mathbf{Q}_{3}}=0.1147\,\mathrm{eV}$. The corresponding HHG spectra normalized by the maximal value is shown in Fig.~\ref{fig:HHG_T_14}. Like the high temperature phase, H3 and H7 also have a single peak, while the H5 splits into two symmetric peaks. Introducing an offset $\alpha=5\text{\textdegree}$ will break the symmetry of HHG spectra (see blue lines in Fig.~\ref{fig:HHG_T_14}). However, this is quite different from the one observed in Ref.~\cite{Tyulnev2024}, where the left peak should be higher than the right peak; see the black lines in Fig.~\ref{fig:HHG_T_14}. Therefore, the asymmetry induced by the crystal axis offset does not explain all of the experimental observations. Introducing the asymmetric band dispersion due to the disorder, residual strain, surface inhomogeneity, etc., could also not explain the experimentally observed low-temperature behavior, as the corresponding effect should be weak from the high-temperature results and lead to similar behaviors both in the high-temperature and low-temperature regime.

To reproduce the low temperature behavior, we note that as the triple-$\mathbf{Q}$ CDW order in $\mathrm{TiSe}_{2}$ arises from three independent real space displacements of Ti and Se atoms along the $\mathbf{Q}_{1,2,3}$ directions, respectively~\cite{Kaneko2018}, there could be weak anisotropy in the CDW order parameters due to the possible strains and disorders in experiments, which leads to different displacements of atoms along these three directions. For the green lines shown in Fig.~\ref{fig:HHG_T_14}, we introduced a slight anisotropy in the CDW order parameters, i.e., $\Delta_{\mathbf{Q}_{1}}=0.1148\,\mathrm{eV}$, $\Delta_{\mathbf{Q}_{2}}=0.1147\,\mathrm{eV}$, and $\Delta_{\mathbf{Q}_{3}}=0.1146\,\mathrm{eV}$, with the effective interaction strength $U_{\mathbf{Q}_{1}}=-1.80965\,\mathrm{eV}$, $U_{\mathbf{Q}_{2}}=-1.80935\,\mathrm{eV}$, and $U_{\mathbf{Q}_{3}}=-1.80905\,\mathrm{eV}$. The shape of the obtained HHG spectra reproduces the behavior observed in Ref.~\cite{Tyulnev2024}. Particularly, the left peak now is higher than the right peak. The center of mass shifts of H5 and H7 to the left and H3 to the right are also reproduced, although slightly overestimated. This suggests that the anisotropic CDW dominates over the offset effect in the low-temperature region, which could explain the experimental observations. We note that here we have chosen the order parameters with $\Delta_{\mathbf{Q}_{1}}>\Delta_{\mathbf{Q}_{3}}$ since $\mathbf{Q}_{1}$ and $\mathbf{Q}_{3}$ are the directions symmetric with respect to the polarization angle $90\text{\textdegree}$ and this choice will lead to a larger band gap in the region with $\theta<90\text{\textdegree}$, thus enhancing the higher-order high harmonic generation process like H5 and H7 and suppressing the lower-order one like H3.

\section{Summary}

In conclusion, we have provided a theoretical explanation for the experimentally observed asymmetric behavior of HHG spectra for TiSe$_{2}$ in Ref.~\cite{Tyulnev2024}. Specifically, we have proposed a simplified phenomenological mean-field model for this material, based on which the HHG spectra are calculated through the time-dependent Schr{\" o}dinger equation. Our model is able to describe the double peak behavior of H5 and the single peak behavior in H3 and H7 as a function of laser polarization angle. For the low temperature phase, we find that the weak anisotropy in the CDW order parameters reproduces the strongly asymmetric behavior of HHG spectra, which could be induced by the possible onset of strain at these temperatures that distorts the CDW phase. This work provides an important step towards establishing high harmonic spectroscopy as a powerful new tool to study the microscopic origins of correlations in materials such as the CDW orders.

\begin{figure}
\includegraphics{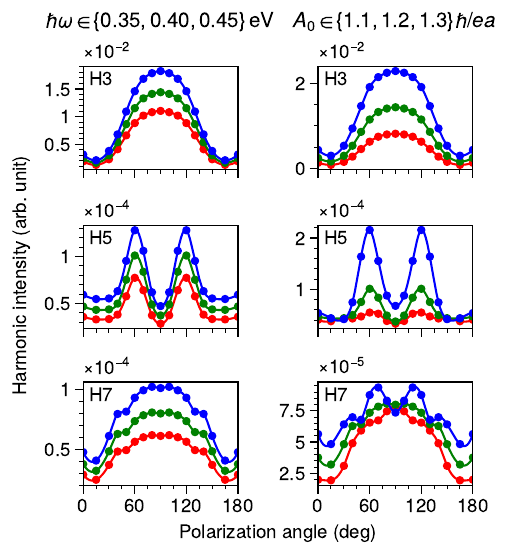}

\caption{HHG spectra of the low temperature CDW phase for various laser parameters in the practical $45\text{\textdegree}$-incident laser geometry. Here the crystal axis offset $\alpha$ is zero, and the CDW order parameters are isotropic, i.e., $\Delta_{\mathbf{Q}_{1}}=\Delta_{\mathbf{Q}_{2}}=\Delta_{\mathbf{Q}_{3}}=0.1147\,\mathrm{eV}$. In the left panel, the different colors represent laser frequency $\hbar\omega=0.35\,\mathrm{eV}$ (red), $0.40\,\mathrm{eV}$ (green), and $0.45\,\mathrm{eV}$ (blue), respectively, while the laser amplitude $A_{0}=1.2\hbar/ea$ and cycle number $n_{\mathrm{cyc}}=8$ are fixed. In the right panel, we considered laser amplitude $A_{0}=1.1\hbar/ea$ (red), $1.2\hbar/ea$ (green), and $1.3\hbar/ea$ (blue), with the fixed laser frequency $\hbar\omega=0.40\,\mathrm{eV}$ and cycle number $n_{\mathrm{cyc}}=8$. \label{fig:stability to laser paramters}}
\end{figure}

\begin{figure}
\includegraphics{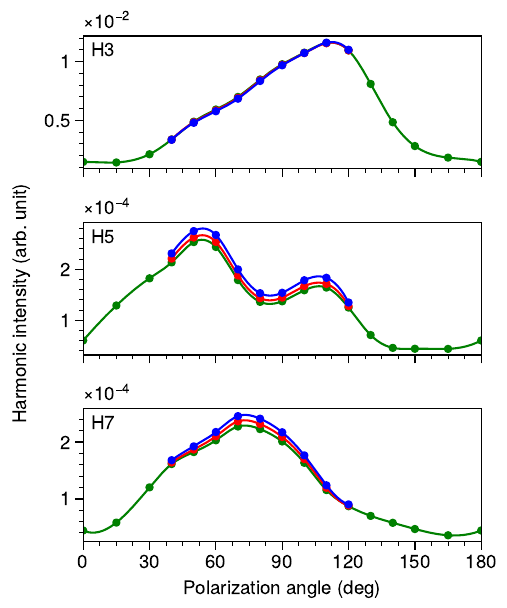}

\caption{HHG spectra of the low temperature CDW phase for various anisotropic CDW order parameters in the practical $45\text{\textdegree}$-incident laser geometry. Here we set $(U_{\mathbf{Q}_{1}}, U_{\mathbf{Q}_{2}}, U_{\mathbf{Q}_{3}})=(-1.80965, -1.80935, -1.80905)\,\mathrm{eV}$ and $(\Delta_{\mathbf{Q}_{1}}, \Delta_{\mathbf{Q}_{2}}, \Delta_{\mathbf{Q}_{3}})=(0.1148, 0.1147, 0.1146)\,\mathrm{eV}$ for the green line; $(U_{\mathbf{Q}_{1}}, U_{\mathbf{Q}_{2}}, U_{\mathbf{Q}_{3}})=(-1.80985, -1.80935, -1.80885)\,\mathrm{eV}$ and $(\Delta_{\mathbf{Q}_{1}}, \Delta_{\mathbf{Q}_{2}}, \Delta_{\mathbf{Q}_{3}})=(0.1149, 0.1147, 0.1145)\,\mathrm{eV}$ for the red line; $(U_{\mathbf{Q}_{1}}, U_{\mathbf{Q}_{2}}, U_{\mathbf{Q}_{3}})=(-1.81035, -1.80935, -1.80835)\,\mathrm{eV}$ and $(\Delta_{\mathbf{Q}_{1}}, \Delta_{\mathbf{Q}_{2}}, \Delta_{\mathbf{Q}_{3}})=(0.1151, 0.1147, 0.1144)\,\mathrm{eV}$ for the blue line. The crystal axis offset is $\alpha=5\text{\textdegree}$, and the laser parameters are the same as in Fig.~\ref{fig:HHG_T_14}. For the latter two cases, we only present the data for the polarization angle $40\text{\textdegree}\leq\theta\leq 120\text{\textdegree}$, from which we can already identify the peak behaviors. \label{fig:stability to CDW anisotropy}}
\end{figure}

\acknowledgments{
The ICFO-QOT group acknowledges support from: European Research Council AdG NOQIA; MCIN/AEI (PGC2018-0910.13039/501100011033, CEX2019-000910-S/10.13039/501100011033, Plan National FIDEUA PID2019-106901GB-I00, Plan National STAMEENA PID2022-139099NB, I00, project funded by MCIN/AEI/10.13039/501100011033 and by the ``European Union NextGenerationEU/PRTR'' (PRTR-C17.I1), FPI); QUANTERA MAQS PCI2019-111828-2; QUANTERA DYNAMITE PCI2022-132919, QuantERA II Programme co-funded by European Union’s Horizon 2020 program under Grant Agreement No 101017733; Ministry for Digital Transformation and of Civil Service of the Spanish Government through the QUANTUM ENIA project call - Quantum Spain project, and by the European Union through the Recovery, Transformation and Resilience Plan - NextGenerationEU within the framework of the Digital Spain 2026 Agenda; Fundaci{\' o} Cellex; Fundaci{\' o} Mir-Puig; Generalitat de Catalunya (European Social Fund FEDER and CERCA program, AGAUR Grant No. 2021 SGR 01452, Quantum CAT\textbackslash{}U16-011424, co-funded by ERDF Operational Program of Catalonia 2014-2020); Barcelona Supercomputing Center MareNostrum (FI-2023-3-0024); Funded by the European Union. Views and opinions expressed are however those of the author(s) only and do not necessarily reflect those of the European Union, European Commission, European Climate, Infrastructure and Environment Executive Agency (CINEA), or any other granting authority. Neither the European Union nor any granting authority can be held responsible for them (HORIZON-CL4-2022-QUANTUM-02-SGA PASQuanS2.1, 101113690, EU Horizon 2020 FET-OPEN OPTOlogic, Grant No 899794, QU-ATTO, 101168628), EU Horizon Europe Program (This project has received funding from the European Union’s Horizon Europe research and innovation program under grant agreement No 101080086 NeQST Grant Agreement 101080086 — NeQST); ICFO Internal ``QuantumGaudi'' project; European Union's Horizon 2020 program under the Marie-Sklodowska-Curie grant agreement No. 847648; ``La Caixa'' Junior Leaders fellowships, ``La Caixa'' Foundation (ID 100010434): CF/BQ/PR23/11980043.
U.B. acknowledges for the financial support of the IBM Quantum Researcher Program. 
R.W.C. acknowledges support from the Polish National Science Centre (NCN) under the Maestro Grant No. DEC-2019/34/A/ST2/00081.
T.G. acknowledges financial support from the Agencia Estatal de Investigaci{\' o}n (AEI) through Proyectos de Generaci{\' o}n de Conocimiento PID2022-142308NA-I00 (EXQUSMI), and that this work has been produced with the support of a 2023 Leonardo Grant for Researchers in Physics, BBVA Foundation. The BBVA Foundation is not responsible for the opinions, comments, and contents included in the project and/or the results derived therefrom, which are the total and absolute responsibility of the authors.
J.B. acknowledges financial support from the European Research Council for ERC Advanced Grant ``TRANSFORMER'' (788218), ERC Proof of Concept Grant ``miniX'' (840010), FET-OPEN ``PETACom'' (829153), FET-OPEN ``OPTOlogic'' (899794), FET-OPEN ``TwistedNano'' (101046424), Lasers4EU (101131771), MINECO for Plan Nacional PID2020–112664 GB-I00; QU-ATTO, 101168628; AGAUR for 2017 SGR 1639, MINECO for ``Severo Ochoa'' (CEX2019-000910-S), Fundaci{\' o} Cellex Barcelona, the CERCA Programme/Generalitat de Catalunya, and the Alexander von Humboldt Foundation for the Friedrich Wilhelm Bessel Prize. I.T. and J.B. acknowledge support from Marie Sk\l{}odowska-Curie ITN ``smart-X'' (860553).
}

\appendix

\section{Tight-binding model of TiSe$_{2}$}

For a full and comprehensive description of our approach, we provide a review of the tight-binding model. It has been
introduced in Ref.~\cite{Kaneko2018}, and gives the highest-lying valence band and the lowest-lying conduction band considered in our phenomenological mean-field model \eqref{eq:mean-field model}. 

The crystal structure of the monolayer TiSe$_{2}$ is given by the primitive translation vectors $\boldsymbol{a}_{1}=(\sqrt{3}a/2,-a/2)$ and $\boldsymbol{a}_{2}=(0,a)$ with lattice constant $a=3.54\,\mathring{\mathrm{A}}$. In each unit cell, we have one Ti ion and two Se ions, Se(1) and Se(2). The corresponding positions are given by $\boldsymbol{\tau}_{\mathrm{Ti}}=(0,0,0)$
and $\boldsymbol{\tau}_{\mathrm{Se1}}=-\boldsymbol{\tau}_{\mathrm{Se}2}=(a/2\sqrt{3},-a/2,z_{\mathrm{Se}})$
with $z_{\mathrm{Se}}=1.552\,\mathring{\mathrm{A}}$, respectively.

The band structure of TiSe$_{2}$ is given by five bands based on the Ti $3d$ orbitals $(d_{x^2-y^2}, d_{3z^2-r^2}, d_{xy}, d_{yz}, d_{zx})$ above the Fermi level and six bands based on the $4p$ orbitals $(p_{x}, p_{y}, p_{z})$ of the Se(1) and Se(2) atoms below the Fermi level. In terms of these orbitals, the tight-binding Hamiltonian is given by 
\begin{equation}
H=\sum_{\mathbf{k}}\sum_{\mu\ell,\nu m}H_{\mu\ell,\nu m}(\mathbf{k})c_{\mathbf{k},\mu\ell}^{\dagger}c_{\mathbf{k},\nu m},
\end{equation}
where $c_{\mathbf{k},\mu\ell}^{(\dagger)}$ is the annihilation (creation) operator of an electron in orbital $\ell$ of atom $\mu$ at momentum $\mathbf{k}$. Here $H_{\mu\ell,\nu m}(\mathbf{k})=\sum_{\boldsymbol{R}_{n}}E_{\mu\ell,\nu m}(\boldsymbol{R}_{n})e^{-\mathrm{i}\mathbf{k}\cdot\boldsymbol{R}_{n}}$ is the Fourier transform of the transfer integral $E_{\mu\ell,\nu m}(\boldsymbol{R}_{n})$ in the Slater-Koster scheme~\cite{Slater1954} between the atomic orbitals $\mu\ell$ and $\nu m$ at coordinate $\boldsymbol{R}_{n}=n_{1}\boldsymbol{a}_{1}+n_{2}\boldsymbol{a}_{2}$ for integers $n_{1}$, $n_{2}$. The energy levels of the atomic orbitals are given by $E_{\mu\ell,\mu\ell}(\boldsymbol{R}_{n}=0)=\varepsilon_{\mu\ell}$. 

Due to the octahedral structure of TiSe$_{2}$, we consider the energy levels $\varepsilon_{d\gamma}$, $\varepsilon_{d\varepsilon}$, and $\varepsilon_{p}$ of the Ti $d\gamma$ ($d_{x^2-y^2}$, $d_{3z^2-r^2}$), Ti $d\varepsilon$ ($d_{xy}$, $d_{yz}$, $d_{zx}$), and Se $p$ ($p_{x}$, $p_{y}$, $p_{z}$) orbitals, respectively. The transfer integrals $E_{\mu\ell, \nu m}(\boldsymbol{R}_{n})$ are obtained through nine transfer integral parameters, $t(pd\sigma)$,
$t(pd\pi)$, $t(dd\sigma)$, $t(dd\pi)$, $t(dd\delta)$, $t(pp\sigma)_{1}$,
$t(pp\pi)_{1}$, $t(pp\sigma)_{2}$, and $t(pp\pi)_{2}$, where $t(pd\sigma)$ and $t(pd\pi)$ are the transfer integrals between the nearest-neighbor Ti $3d$ and Se $4p$ orbitals, and $t(dd\sigma)$, $t(dd\pi)$, and $t(dd\delta)$ are the transfer integrals between the nearest-neighbor Ti-Ti $3d$ orbitals. The subscripts 1 and 2 in $t(pp\sigma)$ and $t(pp\pi)$ indicate the transfer integrals between the nearest-neighbor Se(1)-Se(1) [Se(2)-Se(2)] $4p$ orbitals and between the nearest-neighbor Se(1)-Se(2) $4p$ orbitals, respectively. From Ref.~\cite{Kaneko2018}, the optimized values of the transfer integral parameters for TiSe$_{2}$ are given by $\varepsilon_{d\varepsilon}=0.2\,\mathrm{eV}$, $\varepsilon_{d\gamma}-\varepsilon_{d\varepsilon}=1.112\,\mathrm{eV}$, $\varepsilon_{d\varepsilon}-\varepsilon_{p}=2.171\,\mathrm{eV}$, $t(pd\sigma)=-1.422\,\mathrm{eV}$, $t(pd\pi)=0.797\,\mathrm{eV}$, $t(dd\sigma)=-0.347\,\mathrm{eV}$, $t(dd\pi)=0.119\,\mathrm{eV}$, $t(dd\delta)=-0.030\,\mathrm{eV}$, $t(pp\sigma)_{1}=0.709\,\mathrm{eV}$, $t(pp\pi)_{1}=-0.103\,\mathrm{eV}$, $t(pp\sigma)_{2}=0.592\,\mathrm{eV}$, and $t(pp\pi)_{2}=-0.009\,\mathrm{eV}$.

\section{Stability of our results}

Here we show the stability of our results by considering different laser parameters and CDW anisotropy. Figure~\ref{fig:stability to laser paramters} shows the HHG spectra for various laser parameters with isotropic CDW order parameters and vanishing crystal axis offset. The laser incident angle is $45\text{\textdegree}$. From these plots, we can see that the general peak behavior of HHG spectra is robust across a range of laser parameters. Introducing the nonzero crystal axis offset $\alpha$ and anisotropy in the CDW order parameters will further deform these symmetric HHG spectra into asymmetric one and reproduce the experimentally observed behaviors. In Fig.~\ref{fig:stability to CDW anisotropy}, we also present the HHG spectra for various anisotropic CDW order parameters with fixed laser parameters and crystal axis offset. The asymmetric behavior remains robust. Therefore, our results are stable against laser parameters and CDW anisotropy.

%

\end{document}